\documentclass[a4paper,11pt,final]{article}

\usepackage[english]{babel}
\usepackage[utf8]{inputenc}
\usepackage{graphicx}
\usepackage{algorithm}
\usepackage{algorithmic}
\usepackage{mathtools}
\usepackage{amssymb}
\usepackage{mathptmx}
\usepackage{xcolor}
\usepackage{amsmath}
\usepackage{multirow}
\usepackage[margin=2cm]{geometry}

\usepackage{lastpage}
\usepackage{fancyhdr}
\usepackage{titling}

\usepackage{doi}
\usepackage[autostyle]{csquotes}
\usepackage[style=authoryear,citestyle=authoryear-comp,maxcitenames=2,maxbibnames=99,backend=biber,backref=true,dashed=false]{biblatex}
\usepackage[disable]
	{todonotes}
\newcommand{\todoVL}[1]{\todo[color=red!40, author=\textbf{VL}, inline]{#1}}

\title{Community Structure Characterization}
\author{\vspace{-0.2cm}Vincent Labatut\\
	\vspace{-0.2cm}\small Laboratoire Informatique d'Avignon -- LIA EA 4128, Université d'Avignon, France \\
    \small\href{mailto:vincent.labatut@univ-avignon.fr}{\texttt{vincent.labatut@univ-avignon.fr}} \\
    \vspace{-0.2cm}Günce Keziban Orman\\
	\vspace{-0.2cm}\small Computer Engineering Department, Galatasaray University, 
Ortaköy, Istanbul, Turkey \\
	\vspace{-0.2cm}\small\href{mailto:korman@gsu.edu.tr}{\texttt{korman@gsu.edu.tr}}}

\hypersetup{
    pdftitle={\thetitle},
    bookmarksnumbered=true,bookmarksopen=true,
	unicode=true,colorlinks=true,linktoc=all,
	linkcolor=blue,citecolor=blue,filecolor=blue,urlcolor=blue,
	pdfstartview=FitH
}

\usepackage{enumitem}
\setlist{nolistsep}

\begin{document}
\maketitle

\begin{abstract}
\addcontentsline{toc}{section}{Abstract}
This entry discusses the problem of describing some communities identified in a complex network of interest, in a way allowing to \textit{interpret} them. We suppose the community structure has already been detected through one of the many methods proposed in the literature. The question is then to know how to extract valuable information from this first result, in order to allow human interpretation. This requires subsequent processing, which we describe in the rest of this entry.

\vspace{0.3cm}

\vspace{0.3cm}
\noindent \textcolor{red}{\textbf{Cite as:} V. Labatut \&  G. K. Orman. \href{https://link.springer.com/referenceworkentry/10.1007/978-1-4614-7163-9_110151-1}{Community Structure Characterization}, Encyclopedia of Social Network Analysis and Mining, 2nd Edition, Eds.: R. Alhajj \& J. Rokne, Springer, 2017. Doi: \href{https://doi.org/10.1007/978-1-4614-7163-9_110151-1}{10.1007/978-1-4614-7163-9\_110151-1}}
\end{abstract}

\section{Introduction}
Community detection is one of the most studied problem in the domain of Network Science, as illustrated by the hundreds of algorithms proposed in the literature \parencite{Fortunato2010} and the various definitions of the notion of community itself \parencite{Yang2015e}. However, from the application perspective, detecting the community structure of a network of interest is only half the work. Indeed, this information has no value in itself: one must additionally \textit{interpret} the community structure relatively to the system modeled by the network, in order to bring some sense to it, thus allowing human understanding. Yet, almost all works in the field of community detection deal with the design of detection tools, and the evaluation of their precision or speed \parencite{Fortunato2009}. Very few researchers have addressed the problem of characterizing and interpreting the detected communities \parencite{Tumminello2011, Labatut2012, Labatut2013, Yang2013, Orman2015}. 

In this entry, we consider the interpretation problem as independent from the method used for community detection. We adopt an approach based on the original definition of the notion of community in social sciences, which underlines that nodes belonging to the same community should be relatively similar and/or share a common behavior. Assessing node similarity requires describing nodes, which can be performed both in terms of \textit{individual} information (i.e. personal characteristics) and \textit{relational} information (i.e. connection to the rest of the network). Concretely, the former corresponds to nodal attributes, whereas the latter depends on the network topology. The behavior of a node can be described in terms of evolution of its individual and relational information. This approach allows us to take advantage of most types of information one can encode in a network (structure, directions, weights, attributes, time...).

\section{Key Points}
We review the main methods allowing to characterize communities. We distinguish them depending on the type of information they are based upon: structure only, nodal attributes (which requires an attributed network) and temporal evolution (dynamic networks). Networks encoding more information allow to apply more advanced analysis methods, possibly leading to more informative results. Moreover, interpretation methods can also be distinguished depending on their focus. The community structure is described at the macroscopic scale (that of the whole network), whereas a community can be described at two levels, either by considering it as a whole (mesoscopic scale) or by focusing on its constituting nodes (microscopic scale).

\section{Historical Background}
Community detection as such is quite a recent subfield, dating back to the seminal work by \textcite{Girvan2002}. Related (but different) problems have been the objects of prior works, though, such as graph partitioning or spectral clustering \parencite{Newman2004b}.

In early community detection works, the studied networks were very small, which allowed to interpret the communities manually. In other words, one would study subjectively the individuals composing some community of interest, and try to identify some relevant patterns or regularities in order to reach some observations considered useful to understand the studied system \parencite{Girvan2002,Newman2006}. When the scale of the networks increased from tens to hundreds, it was still possible to consult domain experts to perform interpretation in a similar fashion \parencite{Radicchi2004,Rosvall2007,Rosvall2008}. 

However, this method showed its limit on larger networks. \textcite{Blondel2008} applied their Louvain algorithm on a network of Belgian mobile phone communications, including 2.6 million nodes representing persons. Interpreting such a large network obviously requires a more automatic approach. \citeauthor{Blondel2008} verified the accuracy of the top level of their hierarchical community structure by considering the homogeneity of the nodes on an attribute representing the language people spoke on the phone. This highlights the difficulty of interpreting communities in networks of this size, and also shows one solution can be to consider some additional information, such as nodal attributes.

\section{Notations \& Glossary}
\begin{itemize}
	\item \textbf{Graph or network:} A pair $G=(V,E)$ constituted of a set of nodes $V$ and a set of links $E$. We note $n=|V|$ the number of nodes and $m=|E|$ the number of links.
    \item \textbf{Community structure:} Partition of the node set $V$ into a set of $\lambda$ distinct communities, i.e. $\mathcal{C} = \{C_1,...,C_\lambda\}$, with $V = \bigcup_{i=1}^\lambda C_i$ and $\bigcap_{i=1}^\lambda C_i = \emptyset$.
	\item \textbf{Community:}  Roughly corresponds to a group of nodes more densely interconnected, relatively to the rest of the network. Formally, community number $i$ is a subset of nodes: $C_i \subset V$. We note $n_i$ the number of nodes in $C_i$, and $m_i$ the number of links between these nodes.
    \item \textbf{Attributed network:} Network whose nodes are described by individual attributes, e.g. in a social network: age, gender, ethnicity, etc. 
    \item \textbf{Dynamic network:} Network whose structure and/or attributes evolve through time, represented as a sequence of static consecutive networks. 
    \item \textbf{Time slice:} Static network representing the state of a dynamic network for a given period of time.
\end{itemize}

\section{Focusing on Topology Only}
\label{sec:TopologyOnly}
In some cases, the only available data is the network structure, or alternatively one wants to focus on a purely topological interpretation of the communities. A number of measures exist, which allow to describe one community or the whole community structure. In this entry, we focus on the most widespread ones. They assess the cohesion and separation of the communities, i.e. the way the intra-community links (i.e. links located inside communities) and inter-community links (i.e. links located between communities) are distributed, respectively.

We distinguish two types of measures. The first involves selecting a measure originally designed to describe a whole network, and restricting it to a single community. This generally requires considering the subgraph induced by the community of interest, and processing the measure as originally defined. The second type includes measures specifically designed to characterize communities or community structures.

\subsection{General Measures}
A number of measures have been proposed to characterize complex networks, each one focusing on a specific aspect of their topology (see \parencite{FontouraCosta2007} for a review). The most basic is the \textit{Size} $n_i$, expressed in number of nodes. Studying the size of the detected communities is informative in itself, moreover their distribution can also reveal some properties of the network. Indeed, it was often observed that community size follows a power-law distribution in real-world networks \parencite{Clauset2004}. 

The \textit{Link Density} $\delta$ is a simple measure which can be used to assess community cohesion. It is particularly relevant here, since communities are, by definition, supposed to be more densely connected than the rest of the network. It is defined as: 
\begin{equation}
	\delta(C_i) = \frac{m_i}{n_i(n_i-1)/2}
\end{equation}
i.e. the proportion of existing to possible links inside the community. Certain authors use a normalized version called \textit{Scaled Density} instead: $\delta'(C_i) = n_i\delta(C_i)$ \parencite{Lancichinetti2010}. It has the advantage of taking the value $2$ if the community is a tree,  and $n_i$ if it is a clique.

Cohesion can also be described in terms of \textit{Average Distance} $\ell$: 
\begin{equation}
    \ell(C_i) = \frac{1}{n_i(n_i-1)/2} \sum_{u,v \in C_i} d(u,v)
\end{equation}
where $d(u,v)$ is the geodesic distance between nodes $u$ and $v$. It is worth studying how this measure evolves as a function of the community size, since communities are supposedly small-world \parencite{Lancichinetti2010}.

Another way to characterize cohesion is to use the community transitivity (a.k.a. clustering coefficient) \parencite{Lancichinetti2010}. In its local version, it is defined as:
\begin{equation}
	T(u) = \frac{t_i(u)}{k_{int}(u)(k_{int}(u)-1)/2}
\end{equation}
where $t(u)$ is the number of links between the neighbors of $u$ belonging to its community, whereas $k_{int}(u)$ is the internal degree of node $u$, i.e. its number of neighbors in the same community. The measure corresponds to the proportion of links between the neighbors of $u$, among all possible such connections. A community can be described simply by averaging $T$ over its nodes.

\subsection{Community-Specific Measures}
A simple way to measure the separation of a community structure is to process the proportion of inter-community links $S$ \parencite{Labatut2012}:
\begin{equation}
	S(\mathcal{C}) = 1 - \frac{1}{m} \sum_{i=1}^\lambda m_i
\end{equation}

The internal structure of a community can take various forms, which explains why cohesion can be assessed through several different measures. Certain communities are organized around one or a few hubs, i.e. nodes connected to most of the nodes belonging to the same community, which can have various effects such as a small average distance. This can be assessed through the \textit{Hub-Dominance} measure $h$ \parencite{Lancichinetti2010}: 
\begin{equation}
	h(C_i) = \frac{\displaystyle \max_{u \in C_i}(k_{int}(u))}{n_i - 1}
\end{equation}
The numerator therefore corresponds to the highest degree in the community, when considering only internal links. The measure ranges from $0$ (only isolate nodes) to $1$ (at least one node connected to all others).

The \textit{Embeddedness} $e(u)$ of a node $u$ is a measure assessing both cohesion and separation, since it corresponds to the proportion of the node's neighbors located in the same community:
\begin{equation}
	e(u) = \frac{k_{int}(u)}{k(u)}
\end{equation}
where $k(u)$ is the plain degree of node $u$. If $e(u)$ is close to $1$, the node is particularly well connected to its communities, and \textit{vice versa} if is close to zero. 

The \textit{Within-Community Degree} $z(u)$ is also based on the internal degree, but relies on a $z$-score normalization \parencite{Guimera2005}:
\begin{equation}
	z(u) = \frac{k_{int}(u) - \mu_i(k_{int})}{\sigma(k_{int})}
\end{equation}
where $\mu_i(k_{int})$ is the average internal degree for community $C_i$, and $\sigma(k_{int})$ is its standard deviation. It represents how well a node is connected to its community. It is completed by the participation coefficient $P(u)$ \parencite{Guimera2005}:
\begin{equation}
	P(u) = 1 - \sum_{i=1}^{\lambda}{\left(\frac{k_i(u)}{k(u)}\right)^2}
\end{equation}
where $k_i(u)$ is the community degree of node $u$, i.e. its number of neighbors in community $C_i$. The participation coefficient gets close to $1$ when the node is evenly connected to many communities, and reaches zero when all its neighbors are in the same community. Both measures were originally defined to identify the community roles of nodes, but they also have been used to characterize communities \parencite{Labatut2012}. Some modifications were later proposed to solve certain limitations and generalize them to directed networks \parencite{Dugue2015}.

The quality of the whole community structure can be assessed using one of the many objective functions designed to perform community detection (see \parencite{Fortunato2010} for a very complete review, or more recently \parencite{Creusefond2015}). Among them, the most widespread is clearly Newman's modularity \parencite{Newman2004a}:
\begin{equation}
	Q(\mathcal{C}) = \sum_{i=1}^{\lambda} (\frac{m_i}{m} - \frac{m_{i+}^2}{m^2})
\end{equation}
where $m_{i+}$ is half\footnote{Only \textit{half} for matters of normalization, see note 50 in \parencite{Newman2004a}.} the number of links between $C_i$ and the other communities. The modularity is defined at the node level, so it allows characterizing not only the community structure as a whole, but also individual communities.

A number of measures have been proposed to assess the statistical significance of the estimated community structure (see Section 14 of \parencite{Fortunato2010} for a review). The $B$- and $C$-scores are particularly interesting, because they allow characterizing individual communities (by opposition to the whole community structure) \parencite{Lancichinetti2010a}. They measure, with different levels of precision, the likeliness of observing a community similar to the one at hand, in a random network (using the same null-model than Newman's modularity).

\subsection{Use Examples}
\textcite{Labatut2012} use most of these measures to study a specific network of social relationships between university students. One of their main objective is to characterize individual communities in order to understand the differences between them, and interpret them in the context of the studied system. In other words, it is a case study. As explained later, the authors also have access to nodal attributes, which allows them to complete their purely topological analysis.

The objective of \textcite{Lancichinetti2010} is very different: they do not focus on a single network. Instead, they want to compare different classes of networks (biological, technological, social, etc.), through their community structures. For this purpose, they consider their community size distributions, and study the evolution of several measures describing individual communities (density, average distance, hub dominance, embeddedness), as functions of the community size. They observe some classes of networks are indeed characterized by certain types of community structures.

The work of \textcite{Leskovec2008} also focuses on the community structure as a whole. It builds upon the \textit{Conductance} \parencite{Shi2000}, a graph partitioning objective measure (a normalized cut), to define the notion of \textit{Network Community Profile}. For a given network, they first estimate the maximal conductance for a community of a fixed size. They repeat this process for an increasing community size. Finally, they plot the obtained conductance as a function of the community size. The resulting curve is considered as characteristic of the community structure of the network. \citeauthor{Leskovec2008} show that it discriminates not only between random and real-world networks, but also between several types of real-world networks.

\section{Taking Advantage of Nodal Attributes}
\label{sec:NodalAttributes}
As showed in the previous section, it is possible to characterize the community structure as well as individual communities using only topological information. However, the results are quite limited in terms of the interpretation they allow. 

Fortunately, more and more real-world networks also include non-topological information, taking the form of nodal attributes, i.e. individual information describing each node. Two approaches are possible to take advantage of this information: either consider attributes separately, which can be done through classic statistical tools designed for non-relational data, or consider them jointly with the structural information, which requires using specific tools.

\subsection{Attribute-Only Approaches}
A number of classic statistical tools were designed to characterize groups of objects described by various types of attributes. They can therefore be applied to communities detected by purely topological methods, but whose nodes possess individual attributes. The simplest approach consists in focusing on a single attribute at once. The most straightforward method is to identify, in each community, the most widespread value of the attribute of interest. For instance, \textcite{Labatut2012} study which mobile phones are the most popular in each community of a network of university students. On the same note, \textcite{Palla2007} study the proportion of nodes holding the majority value (community-wise) of the attribute of interest, as a function of the community size.

More advanced statistical tools exist, though. \citeauthor{Labatut2012} propose to study the association between community membership and the attribute of interest. For this purpose, it is possible to test for the significance of this supposed association, using Pearson's chi-square test if the attribute at hand is nominal, or an ANOVA if it is numerical (the community itself being considered as a nominal attribute). Moreover, the strength of the association can be assessed through a collection of measures, among others: Pearson's $\Phi$, Cramér's $V$ and Goodman \& Kruskal's $\lambda$.

It is also possible to take simultaneously several nodal attributes into account. Concerning the statistical tests and association measures presented above to assess the relation between a single attribute and community membership, there exist generalizations allowing to deal with several attributes at once \parencite{Labatut2012}. \citeauthor{Labatut2012} also propose to use discriminant analysis to build models able to predict community membership as a function of several numerical (with Linear Discriminant Analysis) or nominal (with Discriminant Correspondence Analysis) attributes. This results in a set of discriminant factors, whose associated weight represent the discriminant power relatively to the predicted variable (here: the community). The same authors also use Multinomial Logistic Regression for the same purpose. The main limitation of these approaches is the underlying assumption that the most discriminant attributes are the same for all communities. But in fact, any tool able to predict a nominal variable depending on a set of numerical and/or nominal variables could be used instead, including those not making this assumption, such as association rule mining methods.

\textcite{Tumminello2011} precisely focus on the characterization of communities in terms of attribute values, with their approach based on the notion of over-expressed gene. They present their tool as specifically designed to study networks, but it actually ignores the topological information, so it could as well be applied to non-relational data (like the other methods presented in this subsection). In a given community, they consider an attribute value is over-expressed (i.e. characteristic of this community) if it appears among its nodes more often than expected from a null model assuming a hypergeometric distribution.

\subsection{Hybrid Approaches}
By hybrid, we mean that the tools described here consider both the network structure and the node attributes. Indeed, as shown in \parencite{Labatut2013}, when dealing with partitions of the node set, the information conveyed by the network structure can be complementary to that encoded in the node attributes. Therefore, confronting both aspects seems relevant.

The \textit{Homophily} measure (a.k.a. assortativity) is particularly interesting, because of its simplicity and straightforward interpretation. This measure assesses the tendency for nodes to connect with other nodes similar (or dissimilar) to them, relatively to some attribute of interest. Let us consider two paired series constituted of the attribute values of pairs of connected nodes: then the homophily is basically the level of association between these two series. Newman proposes to use Cohen’s Kappa statistic and Pearson’s correlation coefficient for nominal and numeric attributes, respectively \parencite{Newman2003a}. It is generally processed over the whole network, but it can also be used to characterize individual communities, as in \parencite{Labatut2012}, where it is applied to show inter-gender relationships vary among communities in a social network.

On a related note, \textcite{Han2016b} define the \textit{Community Similarity Degree}, a measure originally aiming at describing how homogeneous a community of people is in terms of the interests they share, while taking inter-personal relations into account. \citeauthor{Han2016b} focus on the case of online social networking services, in which users can be connected to other users, and express their interest for certain topics. They want to measure how much users belonging to the same community share these interests. The community similarity degree $Csd(C_i)$ of a community $C_i$ is formally defined as follows:
\begin{equation}
	Csd(C_i) = \frac{r_i/q_i - 1}{n_i - 1}
\end{equation}
where $r_i$ is the total number of manifestations of interest from all users belonging to $C_i$, over all available topics, and $q_i$ is the number of topics for which at least one user belonging to $C_i$ has expressed his interest. So, $r_i/q_i$ can be considered as the average popularity of a topic in $C_i$, expressed in number of manifestations of interest. The ratio of this value to $n_i$ (the number of users in $C_i$) can therefore be interpreted as the average number of manifestations of interest from a user for a topic. The rest of the formula ($-1$ in both the numerator and denominator) is just normalization. The measure ranges from $0$ (no common interest at all between users) to $1$ (all users share exactly the same interests). The measure can be applied to the more general context of attributed graphs, not necessarily representing social networking services. Indeed, a topic can be represented by a nodal attribute, whose value is $1$ if the considered user is interested in this topic, and $0$ otherwise. However, note that these attributes must be \textit{binary}, which can constitute an important constraint, depending on the modeled system.

To study attributed graphs, \citeauthor{Stattner2012} propose a method based on frequent pattern mining, consisting in identifying so-called \textit{Frequent Conceptual Links} \parencite{Stattner2012}. A \textit{Conceptual Link} corresponds to set of links connecting nodes sharing similar attributes. It is considered \textit{frequent} when the size of this link set is above a fixed threshold. This method can be seen as a generalization of the notion of homophily, and was initially used to simplify the network and help understanding it. However, it can also be used to characterize communities, as illustrated in \parencite{Stattner2013}. \citeauthor{Stattner2013} define a set of measures to assess how homogeneous communities are in terms of attributes, and vice versa. The approach is not unlike that adopted in \parencite{Labatut2013} to compare communities and clusters. It is worth noticing that both homophily and frequent conceptual links consider only direct connections between nodes, which can be viewed as a limitation in the sense it has a purely local view.

\textcite{Cai2017} propose a method to jointly detect communities and identify so-called \textit{Community Profiles} in social networking services, by considering jointly the users' relationships, their attributes, the content they produce, and how this content propagates through the network. They explicitly make the assumption of homophily, and define a community as a group of densely connected users sharing similar interests and behaviors. Their method detects a set of topics, each one corresponding to a multinomial distribution over a dictionary (each word has  a certain probability to belong to the topic), and two types of community profiles. The \textit{Content Profile} of a community is a multinomial distribution over the set of topics, reflecting the probability for each topic to be discussed in the community ; whereas its \textit{Diffusion Profile} is a multinomial distribution representing the probability for this community to propagate a certain topic to a given community. Obviously, the latter can be processed only if some sort of information propagation process is taking place on the studied system, and if the information describing it is available.

A number of community detection methods have been specifically designed to use both the network structure and the nodal attributes when identifying the communities (see \parencite{Bothorel2015} for a recent review). It seems natural to suppose some by-product of their processing can be used in some way to ease the interpretation of communities. And indeed, some of them output the most characteristic attributes and/or attribute values of the detected communities (e.g. \parencite{Yang2013}, or \parencite{Cai2017} from the previous paragraph). However, there are two important limitations. First, the overwhelming majority of existing algorithms rely on the (sometimes implicit) assumption of homophily, i.e. communities are supposed to be homogeneous in terms of attributes \parencite{Bothorel2015}. Yet, several experimental works show that this is not necessarily the case in practice, and that the level of homophily can even largely differ from one community to the other in the same network \parencite{Labatut2012,Labatut2013,Stattner2013}. However, certain very recent methods allow heterophily and/or independence, e.g. \parencite{Newman2015a}. Second, it is not always clear which information is used exactly when detecting the communities, especially concerning the network structure. This is due to the fact the problem of community detection is ill-defined, because there is no clear unique definition of what a community is \parencite{Fortunato2016}. Some authors even define the notion of community just in a procedural way, i.e. simply as the output of their community detection method \parencite{Fortunato2010}. It is only recently that certain works tried to propose a typology of the definitions for the concept of community \parencite{Yang2015e}. Moreover, the way attributes and structure are combined to reach some form of compromise is not always clear or controlled. All of this makes it very difficult to characterize the communities based on the outputs of such algorithms.

\section{Considering the Network Evolution}
Besides nodal attributes, time is another aspect that can be used to complement topology-based interpretation methods. Of course, taking advantage of the evolution of the studied system requires both to have access to a proper representation (dynamic network) and to apply an appropriate algorithm (dynamic community detection). There is now a number of methods to perform this task: see \parencite{Aynaud2013a} as well as the entries \textit{Dynamic Community Detection} and \textit{Community Evolution} for a review. However, community detection in dynamic networks is not as widely studied as in static networks, so there are only a few works tackling their characterization. We can distinguish two types of approaches: some works focus purely on how the topology evolves, through the analysis of community events and derived measures, whereas others take the evolution of the nodal attributes into account, sometimes in conjunction with the structure.

\subsection{Structure-Only Methods}
\label{sec:StrucOnlyMeth}
The most direct method to take both structure and time into account is simply to study the evolution of the topological measures presented in Section \ref{sec:TopologyOnly}, e.g. modularity as a function of time \parencite{Kashtan2005}. 

\todoVL{\parencite{Toyoda2003} actually proposed the community events before \citeauthor{Palla2007}. They also proposed a bunch of measures which could be included here.} 
More advanced methods exist though, which are based on the characterization of community evolution through the detection of so-called \textit{Community Events} occurring between two consecutive time slices. \textcite{Palla2007} originally proposed three pairs of opposed events: \textit{Growth} vs. \textit{Contraction} (a community size increases vs. decreases), \textit{Merging} vs. \textit{Splitting} (several distinct communities become one vs. one community gets separated into several ones), and \textit{Birth} vs. \textit{Death} (a community appears vs. disappears). Of course, it is also possible for a community to undergo no event at all. Most authors use these events, or equivalent ones, sometimes under different names, e.g. \textit{Form} for \textit{Birth}, \textit{Dissolve}/\textit{Vanish} for \textit{Death}, \textit{Join}/\textit{Expansion} for \textit{Growth}, \textit{Leave}/\textit{Shrinking} for \textit{Contraction} \parencite{Asur2009, Chen2010e, Greene2010, Brodka2013}. Note that the methods proposed by \citeauthor{Palla2007} based on these events were originally used on overlapping communities, but they also apply to disjoint ones.

The most straightforward use of these events is to count them and study their evolution in order to characterize the community structure, and therefore the network dynamics \parencite{Asur2009, Chen2010e, Greene2010, Brodka2013}. However, they also allow defining the notions of \textit{Community Age}, i.e. the time elapsed since the birth $t_0$ of the community, and \textit{Community Lifetime}, i.e. the time between its birth and disappearance $t_{max}$. Their correlation with other measures can then be studied, for instance \citeauthor{Palla2007} observe community age is correlated with community size in their data: the older the community, the larger it is \parencite{Palla2007}.

\citeauthor{Palla2007} also propose to study some kind of community auto-correlation, by applying Jaccard's coefficient to the node sets of the community of interest considered at two different time slices $t_1$ and $t_2$:
\begin{equation}
	R(C_i,t_1,t_2) = \frac{|C_i(t_1) \cap C_i(t_2)|}{|C_i(t_1) \cup C_i(t_2)|}
\end{equation}
where $C_i(t)$ is the node set of $C_i$ at $t$ and $|C_i(t)|$ is its cardinality. It is then possible to focus on the birth time $t_0$ of the community, and study how $R(C_i,t_0,t)$ evolves as a function of $t$, and/or depending on some other measure such as community size. For instance, in the case of \citeauthor{Palla2007}, the auto-correlation decays faster for larger communities (meaning their members change faster). 

They additionally define the \textit{Stationarity Measure} $\zeta(C_i)$ of community $C_i$ as:
\begin{equation}
	\zeta(C_i) = \frac{\displaystyle \sum_{t=t_0}^{t_{max}-1} R(C_i,t,t+1)}{t_{max} - t_0 - 1}
\end{equation}
where $t_{max}$ is the last time slice before the community disappears. The stationarity can be interpreted as the average proportion of nodes staying in the community at each time slice. \citeauthor{Palla2007} characterize communities by comparing it to their lifetime and size. They observe that, for their data, small stable communities can survive for a long time, whereas small unstable ones have a very short lifespan. The opposite is observed for large communities: stable ones do not last a long time, whereas unstable ones do, because their instability is caused by expansion.

Another set of measures leveraging community events was proposed by \textcite{Asur2009}, among which one is designed to characterize communities. They propose the \textit{Popularity Index}, which aims at assessing the attractiveness of a community, i.e. how much nodes are likely to joint it. The popularity index $Pi(C_i,t)$ of community $C_i$ at time $t$ is:
\begin{equation}
	Pi(C_i,t) = J(C_i,t) - L(C_i,t)
\end{equation}
where $J(C_i,t)$ and $L(C_i,t)$ are the numbers of nodes joining and leaving $C_i$ at time $t$, respectively. On the same note, \textcite{Wang2008c} propose their \textit{Member Stability Measure}, which is a normalized version of $L$. Again, it is worth studying the relation between these measures and other community properties. For instance, \citeauthor{Asur2009} study how $Pi$ correlates with the community size, and observe that for their data, large clusters tend to be more node-attractive \parencite{Asur2009}.

\subsection{Attribute-Based and Hybrid Methods}
As before, the most straightforward way of considering both attributes and time, and possibly also the network structure, is to study the evolution of the measures from Section \ref{sec:NodalAttributes}, e.g. homophily as a function of time. Although we are not aware of any such work, it would also be possible to take advantage of the community events presented in Section \ref{sec:StrucOnlyMeth} to study the evolution of nodal attributes. For instance, by processing the association between the occurrence of a certain type of event and the most widespread attributes in the concerned communities. This would allow answering questions of the types: Do similar (in terms of nodal attributes) communities tend to merge? When communities split, are the resulting smaller communities generally uniform in terms of attributes? Are new-born community uniform?

A more advanced approach consists in using the methods previously designed in the field of data mining for the analysis of natural time series (i.e. not networks), such as the ones reviewed in \parencite{Fu2011}. These are made to extract relevant information from sequences of values of various types (numerical, ordinal or nominal). When focusing only on the node attributes, they can be applied as is, whereas taking structure into account could require to modify the original method. Although the idea of applying/adapting classic data mining methods to networks seems obvious, it is not widespread at all in the domain of community interpretation: to the best of our knowledge, there exist only one work of this type, which we describe briefly here.

\textcite{Orman2015} formalize the community characterization problem as a sequential pattern mining problem \parencite{Mabroukeh2010}. Each node is represented through a sequence of individual descriptor values, describing its evolution through time. A descriptor can directly correspond to a nodal attribute, but it can also be a topological measure: this allows \citeauthor{Orman2015} to represent simultaneously attributes and various aspects of the network topology (degree, centrality, transitivity, etc.). A community is therefore described by the set of sequences representing its nodes. It is characterized by identifying the most relevant sequential patterns among these nodal sequences, which allows detecting common changes in topological features and attribute values over time periods. This relevance is enforced through various constraints. First, to get prevalent patterns, they focus on \textit{frequent} ones (i.e. those supported by a at least a certain number of nodes). Second, to get informative patterns, they only detect \textit{closed} ones, i.e. patterns not included in larger patterns possessing the same support. Third, to get distinctive patterns, they select those with the highest \textit{growth rates}. The growth rate of a pattern is a measure showing how much representative a sequence is inside of a group, relatively to the whole studied population. In this case, a group corresponds to a community. 

The interest of this method is that it allows characterizing each community independently from the others, in terms of both structure and attributes. It also allows detecting outliers, i.e. nodes not following the dominant trend of their community. \citeauthor{Orman2015} apply their method to two real-world networks: DBLP (academic collaborations) and LastFM (music-oriented social media). It results in the extraction of high level information, helping better characterizing and understanding the studied systems. For instance, in DBLP, they associate a scientific domain to the detected communities, and are also able to identify authors on the point of switching to another scientific domain. In LastFM, they focus on the Jazz user group, and can provide a relatively clear interpretation to various communities, such as: users preferring vocal artists, or users only remotely interested in Jazz. However, it is worth noticing the method is likely to produce a large number of sequential patterns, which must then be processed manually for interpretation, resulting in a substantive work for the end-user. But in return, the method does not require making any assumption regarding whether communities would be better characterized by topological information or nodal attributes: the most relevant descriptors will automatically be selected, so the method output can be purely topological or attribute-based, as well as a combination of both.

\section{Key Applications}
The tools presented in this entry aim at characterizing communities and community structures, in order to ease their interpretation by human operators. Community detection is itself a very general analysis, which can be performed on any complex network, whatever the modeled system. Therefore, the presented tools are likely to be used on any system.

More precisely, some of them are designed to describe the community structure as a whole, which is useful to compare graphs at a mesoscopic scale, whether they represent distinct systems or a given system at different times. Others characterize communities individually, which is more appropriate to focus on specific communities of interest, understand them, and compare them. 

Articles mentioned in this entry include applications to real-world social networks \parencite{Labatut2012, Labatut2013, Stattner2012}, social media \parencite{Leskovec2008, Lancichinetti2010, Orman2015, Dugue2015}, the Internet \parencite{Guimera2005, Lancichinetti2010}, the Web (or parts of it) \parencite{Kashtan2005, Leskovec2008, Asur2009, Lancichinetti2010}, biological networks \parencite{Guimera2005, Chen2010e, Lancichinetti2010}, communication networks \parencite{Palla2007, Chen2010e, Greene2010, Lancichinetti2010, Brodka2013}, collaboration networks \parencite{Guimera2005, Palla2007, Asur2009, Tumminello2011, Orman2015}, transportation networks \parencite{Guimera2005}, sale co-occurrence networks \parencite{Stattner2013}, electronic circuits \parencite{Kashtan2005}.

\section{Future Directions}
There are now countless community detection methods, but the need for a way to extract meaningful information from the detected communities is still very strong. The future works in this field could follow two complementary ways. First, certain existing tools need improvement in terms of reliability, usability (noticeably parameter estimation, e.g. thresholds), computational complexity, and quality of the produced results. Second, the methods presented here do not allow taking into account all the possible information one can encode in a network: it is necessary to extend them, or propose new ones, to deal with multilayer/multiplex networks, as well as signed relationships (in conjunction with attributes, and temporal evolution). It is worth noticing data mining researchers deal with very similar problems, but applied to non-relational data (i.e. not  networks). Their tools would therefore constitute a very relevant base in the constitution of new, network-related methods, but this source of inspiration has been largely ignored until now by the complex network scientists.

\section*{Acknowledgments}
\addcontentsline{toc}{section}{Acknowledgments}
This article is supported by the Galatasaray University Research Fund (BAP) within the scope of the project number 14.401.002, and titled "Sosyal Ağlarda Küme Bulma ve Anlamlandırma: Zamana Bağlı Sıralı Örüntü Uygulaması".

\renewcommand*{\bibfont}{\footnotesize}
\addcontentsline{toc}{section}{\refname}
\printbibliography

\end{document}